\documentclass[11pt]{article}

\usepackage[
    letterpaper,
    margin=1in
]{geometry}

\usepackage[T1]{fontenc}
\usepackage{amsmath,amssymb,amsfonts}
\usepackage{newtxtext,newtxmath}
\usepackage{pdflscape}

\usepackage{microtype}

\usepackage{graphicx}
\graphicspath{{media/}}

\usepackage{booktabs}
\usepackage{threeparttable}

\usepackage{caption}
\usepackage{enumitem}
\setlist{nosep}

\usepackage{titlesec}

\titleformat{\section}
    {\large\bfseries}
    {\thesection}
    {0.6em}
    {}

\titleformat{\subsection}
    {\normalsize\bfseries}
    {\thesubsection}
    {0.6em}
    {}

\titleformat{\subsubsection}
    {\normalsize\bfseries}
    {\thesubsubsection}
    {0.6em}
    {}

\titlespacing*{\section}
    {0pt}
    {1.8ex plus .4ex minus .2ex}
    {0.8ex}

\titlespacing*{\subsection}
    {0pt}
    {1.4ex plus .3ex minus .2ex}
    {0.5ex}

\titlespacing*{\subsubsection}
    {0pt}
    {1.2ex plus .3ex minus .2ex}
    {0.4ex}

\makeatletter
\renewcommand{\maketitle}{%
  \thispagestyle{plain}
  \begin{center}
    \vspace*{1em}
    \hrule height 1.2pt
    \vspace{1.1em}

    {\LARGE\bfseries \@title \par}
    \vspace{1.25em}

    {\large
      \def\and{%
        \end{tabular}\hfill\linebreak[0]\hfill%
        \begin{tabular}[t]{c}}
      \begin{tabular}[t]{c}
        \@author
      \end{tabular}\par}
    \vspace{1em}

    {\normalsize \@date \par}
    \vspace{1.1em}

    \hrule height 1.2pt
  \end{center}
  \vspace{1.5em}
}
\makeatother

\usepackage[authoryear,round]{natbib}
\usepackage{fancyhdr}

\newcommand{\shortauthors}{Ferraz \& Houf}
\newcommand{\shorttitle}{Opposite effects of human and optimising-machine labels on escalation in a live auction}

\fancypagestyle{plain}{
    \fancyhf{}
    \fancyfoot[C]{\thepage}
    
}

\usepackage[hidelinks]{hyperref}
\title{
    Opposite effects of human and optimising-machine labels on escalation in a live auction
}

\author{
    Vinicius Ferraz \\
    Karlsruhe Institute of Technology (KIT) \\ \& Singularity AI
    \and
    Leon Houf \\
    Karlsruhe Institute of Technology (KIT) \\
    Leon.Houf@kit.edu
}

\date{ }

\begin{document}

\maketitle

\begin{abstract}
People increasingly compete against AI agents, and organisations decide what to tell them about these opponents. We distinguish an \textit{opponent effect}, specific to the system faced, from an \textit{information effect}, a property of the disclosure. In a preregistered, deception-free experiment (N = 1,395), we crossed what participants were told about their opponent with whom they faced in a live all-pay auction in which escalation arises from the incentives. Holding the opponent fixed, being told a human was or might be the opponent raised the median price by 6.7\% of the prize value, and being told an optimising machine might be lowered it by 8.8\%. Facing an AI agent lowered prices and made it rarer that both players finished with positive earnings. Exploratory analyses indicate that the labels changed how often people raised again while already ahead, not how often they responded when behind. Describing an AI competitor is not behaviourally neutral.
\end{abstract}

\section*{Introduction}

Algorithms increasingly act as competitors rather than as tools. They set prices against human rivals \citep{calvano2020artificial, assad2024algorithmic, brown2023competition} and bid against human advertisers \citep{decarolis2021mad}, and the people competing with them are often told little about what they face. Whether and how they are told is becoming a regulatory choice. The European Union's AI Act requires that people be informed when they interact with an AI system \citep{euaiact2024}, but leaves open what they should be told. This makes it a practical question whether the description of an artificial competitor changes how people compete.

A person who learns that the opponent is, or might be, a machine can change their behaviour for two reasons. The machine may play differently from a person, in which case the change belongs to that particular system and need not carry over to another; we call this an \textit{opponent effect}. Alternatively, the description alone may change behaviour, in which case the change belongs to the disclosure, which an organisation chooses regardless of the system behind it; we call this an \textit{information effect}. The second is a framing effect in the sense of \cite{liberman2004name}: a label changes what players expect of each other without changing the game. The two effects call for different responses. An opponent effect concerns how a system is built, an information effect how it is described.

Research on how people respond to machines has mostly studied them as advisors \citep{onkal2009relative, yeomans2019making, longoni2019resistance, Althaus2026Chatbot}, where people sometimes discount and sometimes favour algorithmic judgement \citep{dietvorst2015algorithm, logg2019algorithm}, depending on the task and on how much control they retain \citep{castelo2019task, burton2020systematic, ferraz2025trust}. As counterparts, machines have been studied mainly in cooperation. They can sustain more cooperation than people, but only while their identity is hidden \citep{ishowo2019behavioural}, and they foster cooperation when built for it \citep{crandall2018cooperating, shirado2017locally}. Competition against machines has received less attention \citep{march2021strategic, chugunova2022we}, and studies typically vary either what participants are told or whom they face, rarely both. \cite{teubner2015impact} held opponent behaviour fixed by replaying human bids under a computer label, in a static sealed-bid auction in which escalation cannot develop. Separating the two channels requires crossing the description with the actual opponent, and doing so without deception \citep{hertwig2001experimental, gneezy2026experimental} requires that every statement participants read is true.

Competition is also where a well-documented human failure operates: escalation of commitment, the tendency to keep investing in a contest past the point at which it pays \citep{staw1976knee, brockner1992escalation, sleesman2012cleaning}. Its known determinants lie mainly with the decision-maker and the decision \citep{whyte1986escalating, sweis2018sensitivity}, and whether information about the opponent is among them is open. The dollar auction makes the logic concrete \citep{shubik1971dollar, o1986international, leininger1989escalation}. Once both players have bid, conceding forfeits what one has already paid, so each keeps raising, and the bidding climbs past the value of the prize. Competitive arousal is known to drive such bidding in live auctions \citep{ku2005towards, adam2015auction}.

We separate the two channels in a preregistered experiment with 1,395 participants in 1,033 pairs recruited on Prolific \citep{palan2018prolific}. Pairs played 30 rounds of a dynamic all-pay auction, a repeated war of attrition in which both bidders pay their own bid and only the higher bid wins a prize of 100 points \citep{bulow1999generalized, krishna1997analysis, dechenaux2015survey}, so that escalation arises from the incentives. Participants were told the type of their opponent for certain, told that it was one of two types with equal probability, or told nothing; the types were a human, an AI trained to imitate humans and an AI trained to play optimally. They then faced an opponent consistent with what they had read. Holding the opponent fixed, being told that a human was or might be the opponent raised the median price by 6.7\% of the prize value, and being told that an optimising machine might be lowered it by 8.8\%, a spread of about 15\% produced by information alone. A generic AI label and the imitating-AI label had no detectable effect. Facing an AI agent also lowered prices, but made it rarer that both players finished with positive earnings, 5.1\% of pairs compared with 14.9\% of human pairs. Exploratory analyses indicate that the labels changed how often participants raised again while already ahead, not how often they responded when behind, and that the information effect does not require believing the opponent was human: most participants guessed they had faced an AI.

\section*{A crossed, deception-free design separates what people are told from whom they face}

The design separates what a person is told about the opponent from how the opponent actually behaves. It crosses the two dimensions deception-free. What participants are told in the instructions differs in two ways. It differs in \textit{certainty}, naming the opponent outright, giving an even chance between two types, or, in the no-information condition, saying only that they face ``an opponent". And it differs in how an \textit{AI is characterised}, described in some conditions by its training objective, to imitate people or to play optimally against them, and in others as ``an AI" with no further detail. Who participants actually face is set separately: another human, an AI trained to imitate human bidding (AI-Imitate), or an AI trained to play optimal against humans (AI-Optimal). 
Each treatment ran as its own session, and in sessions naming an even chance between two types the opponent was drawn at random. Because descriptions and opponents are crossed across treatments, each effect can be estimated net of the other, and because every statement is true, a description is paired only with an opponent it fits  
The empty combinations do not prevent identification, because each opponent type occurs under several descriptions.
Table \ref{tab:design} gives the full set of conditions with the number of participants in each.

\begin{table}[tbhp]
\centering
\begin{threeparttable}
\caption{Experimental design. What participants are told (rows) is crossed with the opponent they actually face (columns); cells give numbers of pairs. }
\label{tab:design}
\begin{tabular}{@{}lcccc@{}}
\toprule
 & \multicolumn{3}{c}{\textbf{Actual opponent}} & \\
\cmidrule(lr){2-4}
\textbf{What participants are told} & Human & AI-Imitate & AI-Optimal & Total \\
\midrule
\textit{No information} & & & & \\
\quad ``an opponent''                        & 49  & 56  & 55  & 160 \\[3pt]
\textit{Certain (one opponent named)} & & & & \\
\quad ``a human''                            & 206 & --- & --- & 206 \\
\quad ``AI trained to imitate humans''    & --- & 52  & --- & 52  \\
\quad ``AI trained to play optimally''    & --- & --- & 55  & 55  \\[3pt]
\textit{Uncertain, AI described (50/50)} & & & & \\
\quad ``human or AI-Imitate''                & 28  & 59  & --- & 87 \\
\quad ``human or AI-Optimal''                & 27  & --- & 59  & 86 \\
\quad ``AI-Imitate or AI-Optimal''           & --- & 54  & 55  & 109 \\[3pt]
\textit{Uncertain, AI generic (50/50)} & & & & \\
\quad ``human or AI''                        & 52 & 60  & 58  & 170 \\[3pt]
\textit{AI only, generic} & & & & \\
\quad ``an AI''                              & --- & 55  & 53  & 108 \\
\midrule
\textbf{Total Pairs}                               & 362 & 336 & 335 & \textbf{1{,}033} \\
\bottomrule
\end{tabular}
\begin{tablenotes}[flushleft]
\footnotesize
\item \textit{Notes:} Human--human pairs contain two human participants; human-AI pairs contain one, so 1,033 pairs comprise 1,395 participants. ``AI-Imitate" and ``AI-Optimal" are the short forms for ``AI trained to...". Participants always saw the full descriptions. Cells report numbers of pairs. Because every statement is
truthful, a description is paired only with an opponent consistent with it, so
combinations that would require deception are empty. 
\end{tablenotes}
\end{threeparttable}
\end{table}

Escalation is measured in a dynamic all-pay auction, a real-time war of attrition \citep{smith1974theory, bulow1999generalized, krishna1997analysis} in which the runaway bidding arises from the incentives themselves rather than from anything we instruct. Two players compete for a fixed prize of 100 points by raising a shared standing bid in steps of 5, following the structure of the dollar auction \citep{shubik1971dollar}, each player bidding in real time. When the round ends, both players pay their final bid, but only the higher bidder wins the prize, and only the player currently behind can choose to stop. A round also ends if that player neither bids nor concedes within 10 seconds. This incentive mechanism creates a trap. Suppose the leading bid is 100 and a participant's own bid is 95: conceding means losing 95 for nothing, whereas bidding 105 and winning means losing only 5. Continuing beats conceding even though it pushes the price above the prize. The opponent, once behind, faces the identical choice, and so the bidding climbs past the point where either can profit. Pairs play 30 such rounds with earnings accumulating across them, so the design records both how far bidding escalates and how it unfolds over time. The auction runs in the browser in real time, implemented in oTree
\citep{chen2016otree}, with bot defence measures implemented \citep{rilla2026recognising}. 

Two artificial opponents make the AI descriptions truthful, one trained to imitate people and one trained to compete well against them. Both share an architecture trained on approximately 104,000 actions from 206 human pairs: one neural-network head chooses whether to continue or concede, a second reproduces human-like response times so the agents do not act faster than a person would. The imitating agent is trained by behaviour cloning to match typical human play, an approach used to build agents that reproduce human decisions rather than optimal ones \citep{mcilroy2020aligning, meta2022human}. The optimising agent uses the same data but weights each decision by how well the round turned out for that player, favouring the more successful choices \citep{peng2019advantage}, and is regularised toward the imitating agent so that it stays behaviourally plausible rather than drifting into inhuman play. ``Optimal" refers to this objective and method, not to any equilibrium guarantee, which a repeated dynamic all-pay auction does not admit; the instruction participants read is therefore accurate on its own terms. Both agents use fixed parameters and do not learn from the participant. Methods and Supplementary Information give full detail.

The condition in which participants are told nothing and face a human anchors the design: it is the reference against which both effects are measured. For each pair~$i$ we model the outcome as
\begin{equation}
  Y_i = \alpha + \boldsymbol{\beta}'\,\mathbf{Actual}_i
      + \boldsymbol{\gamma}'\,\mathbf{Info}_i + \varepsilon_i,
  \label{eq:main}
\end{equation}
where $\mathbf{Actual}_i$ indicates the opponent actually faced
(AI-Imitate or AI-Optimal, human omitted) and $\mathbf{Info}_i$
indicates each opponent a participant was told was possible (no-information omitted). 
An indicator is active whenever its opponent type was named as possible, whether for certain or with an even chance; a participant told ``human or AI-Imitate'' therefore has two indicators active, and one told only ``an opponent'' has none.
The omitted categories make the intercept $\alpha$ the reference cell; the coefficients $\boldsymbol{\gamma}$ then
measure the information effect and $\boldsymbol{\beta}$ the opponent
effect, each net of the other. Because information and opponent are
crossed, the two are separately identified:
$\boldsymbol{\gamma}$ is estimated free of differences in who is faced,
and $\boldsymbol{\beta}$ free of differences in what is told. This
specification is common to all outcome variables, with additional controls where stated. For the preregistered, primary outcome, average pair-level price over the 30 rounds, we estimate at the median (quantile regression, $\tau = 0.5$) rather than the mean, because a minority of pairs escalate into a long right tail that would otherwise dominate an average and misrepresent the typical pair. Confidence intervals throughout are the estimate $\pm$ 1.96 standard errors from a wild bootstrap with 2,000 draws, clustered by pair where a pair contributes several observations (Methods).

\section*{Results}

\subsection*{Human and optimising-machine labels move prices in opposite directions}

\begin{figure}[!htbp]
  \begin{minipage}[t]{0.49\linewidth}
    {\sffamily\bfseries\fontsize{8}{8}\selectfont a}\\[0.5mm]
    \includegraphics[width=\linewidth]{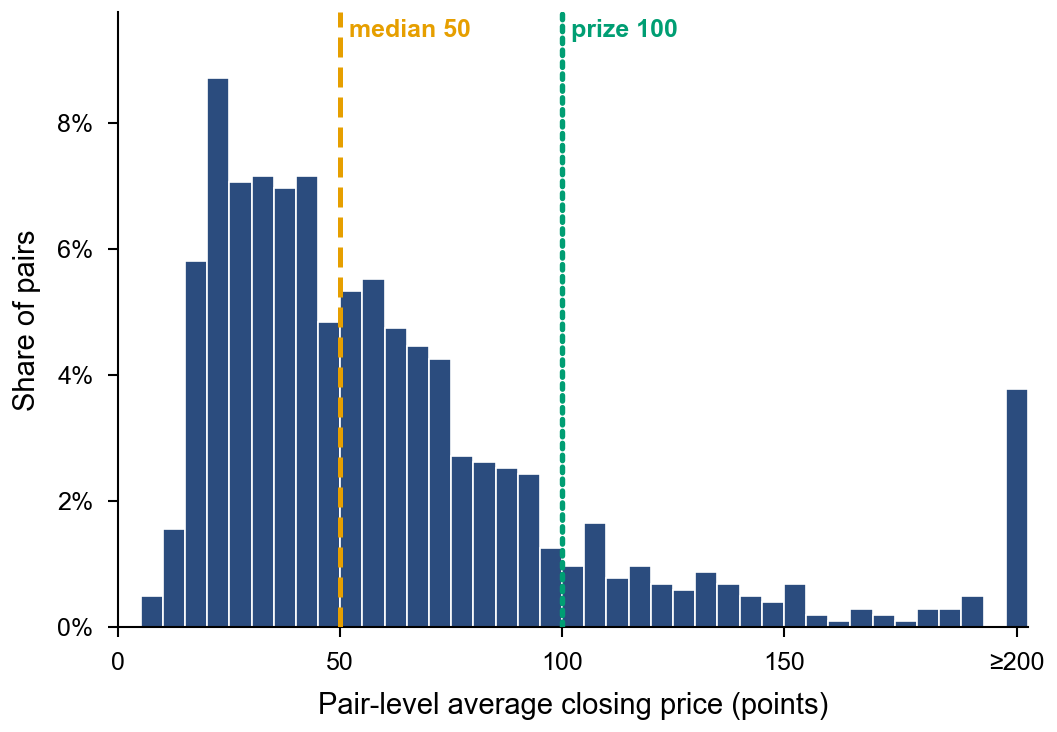}
  \end{minipage}\hfill
  \begin{minipage}[t]{0.49\linewidth}
    {\sffamily\bfseries\fontsize{8}{8}\selectfont b}\\[0.5mm]
    \includegraphics[width=\linewidth]{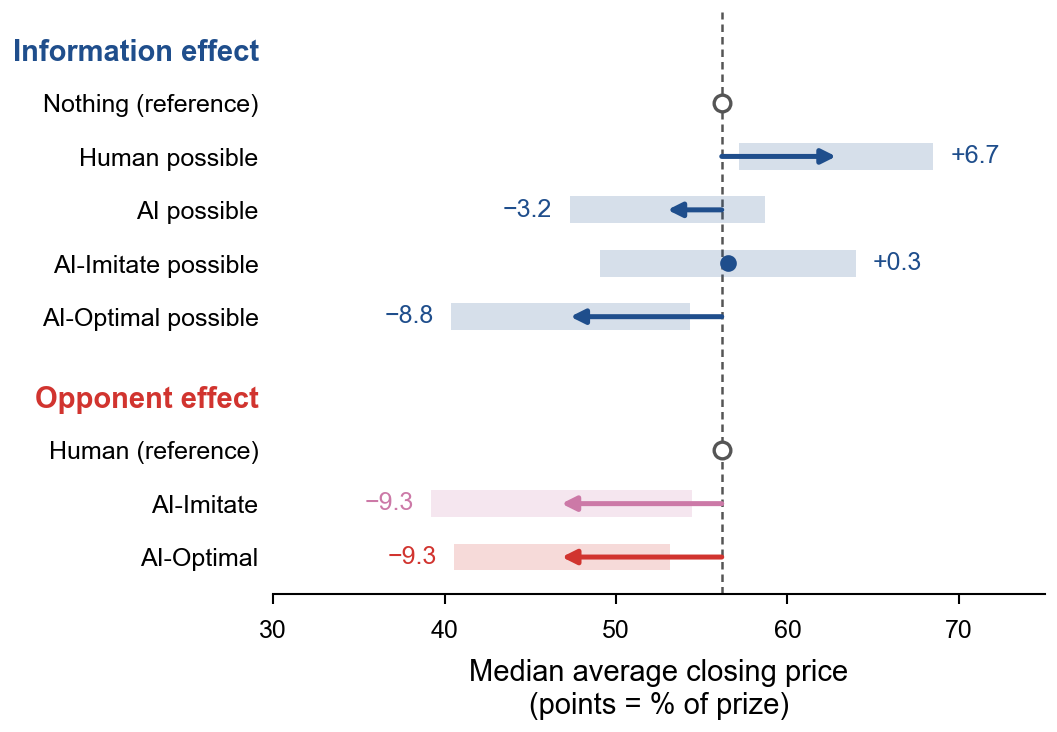}
  \end{minipage}\\[3mm]
  \begin{minipage}[t]{0.49\linewidth}
    {\sffamily\bfseries\fontsize{8}{8}\selectfont c}\\[0.5mm]
    \includegraphics[width=\linewidth]{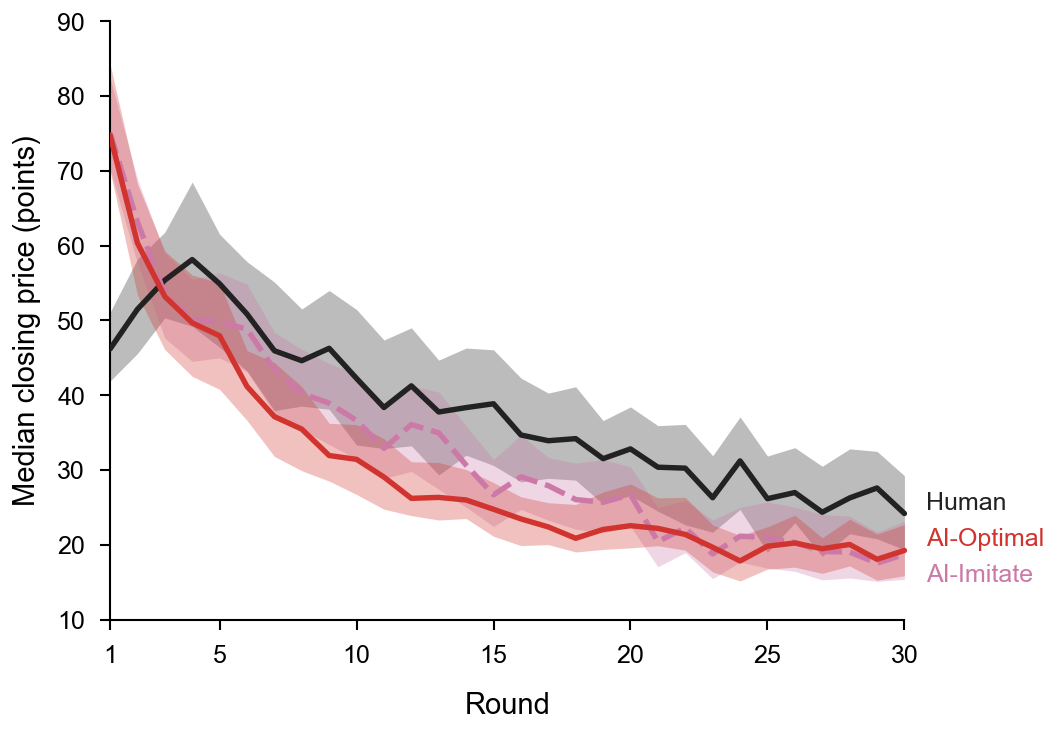}
  \end{minipage}\hfill
  \begin{minipage}[t]{0.49\linewidth}
    {\sffamily\bfseries\fontsize{8}{8}\selectfont d}\\[0.5mm]
    \includegraphics[width=\linewidth]{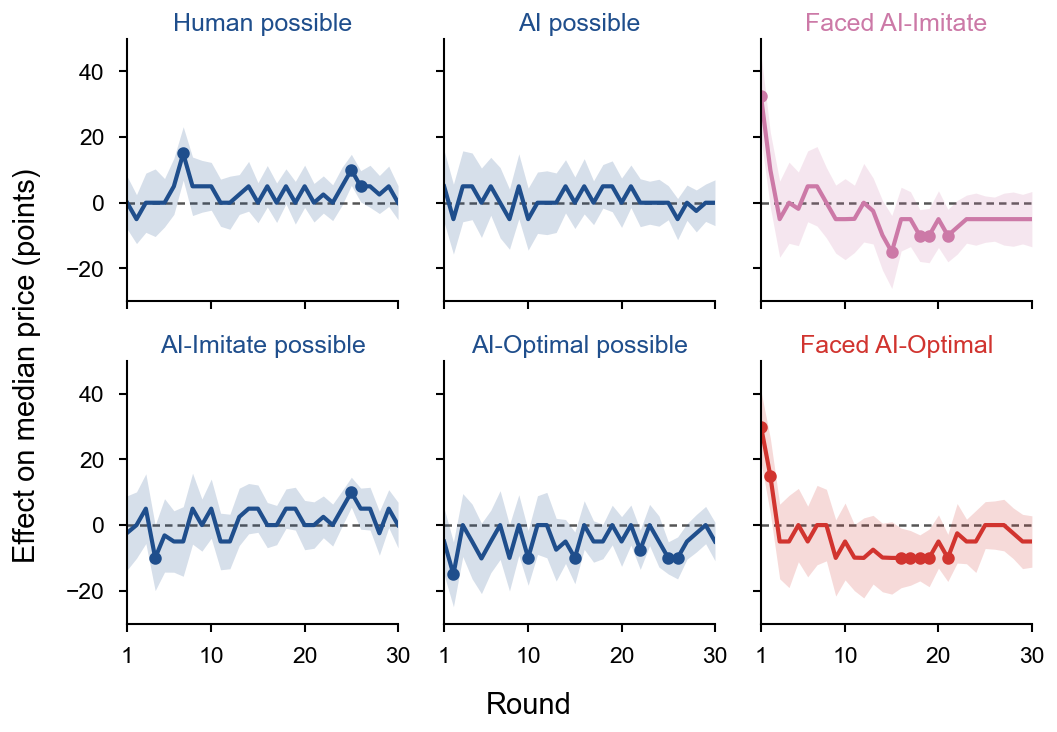}
  \end{minipage}
  \caption{\textbf{ Human and optimising-machine labels push prices in opposite directions, separately from the opponent faced.}
    \textbf{a}~Distribution of the pair-level average closing price over 30 rounds across all conditions ($n = 1{,}033$ pairs). The dashed line marks the median (50 points) and the dotted line the prize value (100 points). Values of 200 or more are pooled in the last bar; 14.2\% of pairs average more than the prize and 3.8\% more than twice the prize.
    \textbf{b}~Effects from the preregistered median regression ($\tau = 0.5$) of the pair-level average closing price on indicators for each opponent participants were told was possible (blue) and for the opponent actually faced (red and pink), estimated jointly. Arrows start at the estimated median of the reference cell (told nothing, faced a human; 56.2 points, 95\% CI [49.8, 62.6]; dashed line) and end at the predicted median with one indicator switched on. Shaded bars show the 95\% confidence interval of each effect; a bar crossing the dashed line marks an effect not distinguishable from zero.
    \textbf{c}~Median closing price per round by opponent actually faced (human, $n = 362$ pairs; AI-Imitate, $n = 336$; AI-Optimal, $n = 335$). Shaded bands show 95\% percentile intervals from a bootstrap that resamples pairs with their full 30-round trajectories.
    \textbf{d}~Coefficients from the median regression in Eq.~(1) estimated separately for each round, one observation per pair, on a common y-axis. Shaded bands show 95\% confidence intervals; dots mark rounds with $p < 0.05$ (unadjusted). Per-round estimates are exploratory.
    Confidence intervals in \textbf{b} and \textbf{d} are the estimate $\pm$ 1.96 bootstrap standard errors from a wild bootstrap with Rademacher weights (Feng et al., 2011). All bootstrap procedures use 2,000 draws. AI-Imitate and AI-Optimal abbreviate the descriptions participants read (``AI trained to imitate humans'' and ``AI trained to play optimally''). Source data are provided as a Source Data file.}

  \label{fig:price}
\end{figure}

Escalation is common. The median pair-level average closing price is 50 points, half the prize, but the distribution has a long right tail, as is typical of contest experiments \citep{sheremeta2013overbidding, dechenaux2015survey, kirchkamp2021spite}: 14.2\% of pairs average more than the prize over their 30 rounds, and 3.8\% average 200 or more (Fig.~\ref{fig:price}a). Because a few runaway pairs would dominate an average, we estimate effects at the median, as preregistered, so that they describe the typical pair.

What participants are told about the opponent changes how far they escalate (Fig.~\ref{fig:price}b). In the reference cell, where participants were told nothing and faced a human, the median price is 56.2 (95\% CI [$49.8$, $62.6$]). Holding the opponent fixed, being told that a human was or might be the opponent raises the median price by $6.7$ points, i.e., $6.7\%$ of the prize value (95\% CI [1.0, 12.3], $p = 0.021$), and being told that an optimising machine might be lowers it by 8.8\% ([$-15.8$, $-1.8$], $p = 0.013$). The two descriptions differ by 15.5\% of the prize value ([$5.9$, $25.1$], $p =0.002$), a gap produced by the description alone.

The effect is specific to these two descriptions rather than to the distinction between a human and an AI. A generic ``an AI'' label and the description of an AI trained to imitate humans have no detectable effect ($-3.2$, $p = 0.28$; $+0.3$, $p = 0.93$). Only the prospect of a person and of a machine described as playing optimally move behaviour, and they move it in opposite directions. Because every estimate holds the actual opponent fixed, these effects are not a by-product of how our agents play.

\subsection*{Facing an agent lowers prices and rarely leaves both players ahead}

Who participants face also matters. Facing either agent lowers the median price by 9.3\% of the prize value relative to a human opponent (AI-Imitate: 95\% CI [$-17.0$, $-1.7$], $p = 0.016$; AI-Optimal: [$-15.6$, $-3.0$], $p = 0.004$), and the two agents produce almost identical reductions (Fig.~\ref{fig:price}b). The difference emerges over the first rounds (Fig.~\ref{fig:price}c). Pairs with an agent start higher, with a median closing price of about 75 in the first round against about 47 between two people, but fall below human pairs by the third or fourth round and stay there, while prices decline over the 30 rounds in every group. Estimating Eq.~(\ref{eq:main}) separately for each round shows the same reversal as a large positive opponent coefficient in the first round followed by negative ones (Fig.~\ref{fig:price}d). The information effects vary from round to round without a clear pattern, and we treat these per-round estimates as exploratory.

The agents also change a second, preregistered outcome: whether both players finish the 30 rounds with positive earnings, so that both receive a bonus. Such mutual success is uncommon even between two people, at 14.9\% of human pairs, and rarer when an agent is involved, at 5.1\%. In a pair-level logistic regression on the indicators of Eq.~(\ref{eq:main}), facing AI-Imitate lowers the probability of mutual success by 8.1 percentage points (95\% CI [$-12.6$, $-3.5$], $p < 0.001$) and facing AI-Optimal by 6.2 percentage points ([$-10.9$, $-1.5$], $p = 0.010$). Of the descriptions, only the optimising-machine label is associated with lower mutual success ($-4.6$ percentage points, [$-8.9$, $-0.3$], $p = 0.035$). Two people sometimes settle into an implicit truce, alternating wins at low prices as symmetric repeated games tend to elicit \citep{kaplan2012way}; pairs with an agent seldom do.

This is a property of the agents we built rather than of machine opponents in general. Machines sustain cooperation when they are built for it \citep{crandall2018cooperating}, and ours were not. Both learned from 206 human pairs, of which only 28 reached mutual success, too few for either agent to learn the coordination behind it. The opponent effect therefore describes these agents, whereas the information effect is estimated net of them.

\subsection*{The labels change how often people raise again while ahead, not how often they respond when behind}

Each bid raises the price by 5 and is one of two kinds. A response follows the opponent's bid, when the bidder is behind; a re-raise follows the bidder's own bid, when they already lead and raise their own bid nonetheless. Re-raising is common. For the typical participant, re-raises account for a little over half of the price they add through their own bids (median 54\%; 65\% pooled over all bids). Around the prize the share of re-raises dips, so that the prize is crossed mainly by responding, and beyond it re-raises become more frequent again.

\begin{figure}[!htbp]
  \begin{minipage}[t]{0.99\linewidth}
    {\sffamily\bfseries\fontsize{8}{8}\selectfont a}\\[0.5mm]
    \includegraphics[width=\linewidth]{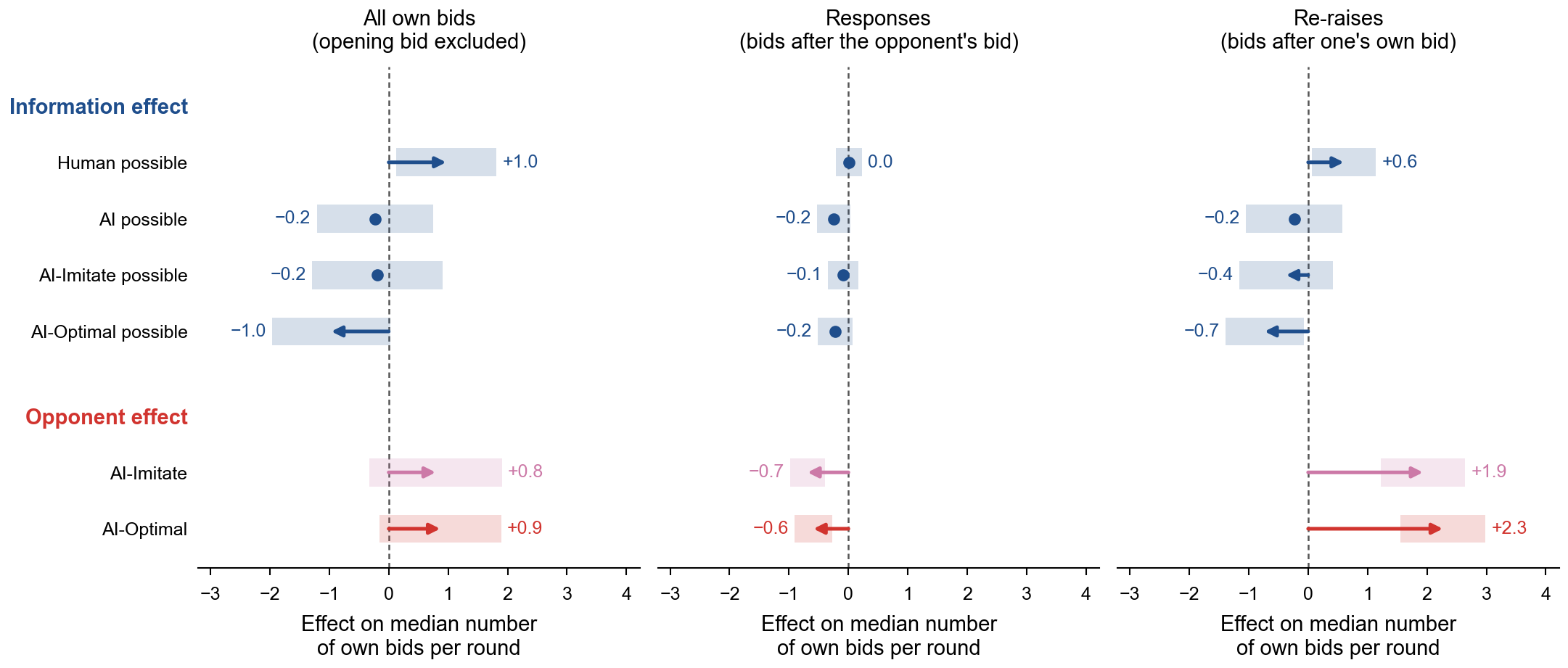}
  \end{minipage}\hfill
  \\[3mm]
  \begin{minipage}[t]{0.99\linewidth}
    {\sffamily\bfseries\fontsize{8}{8}\selectfont b}\\[0.5mm]
    \includegraphics[width=\linewidth]{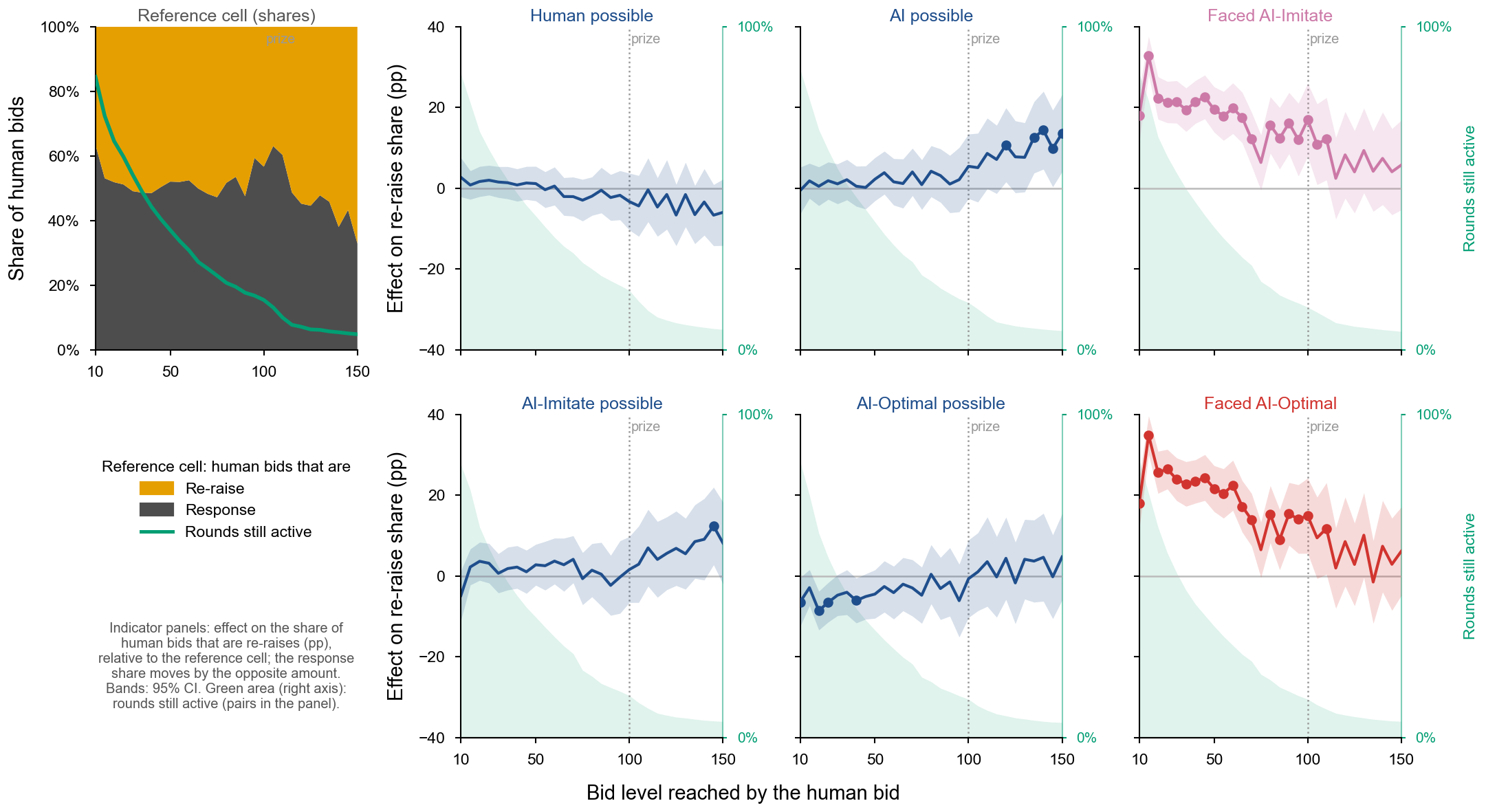}
  \end{minipage}\hfill
\caption{\textbf{What participants are told changes how often they raise again while already ahead, and facing an agent shifts their bids from responses to re-raises.}
\textbf{a} Effects on the number of each participant's own bids per round, averaged over 30 rounds, from median regressions ($\tau = 0.5$) on the indicators of Eq.~(1) ($n = 1{,}395$ participants in 1,033 pairs). The opening bid of each round is excluded and rounds without a bid count as zero; each bid raises the price by 5. The left column covers all own bids; the other two split them into responses (bids after the opponent's bid) and re-raises (bids after one's own bid). Medians do not add up, so the two components need not sum to the total. Estimates are relative to the reference cell (told nothing, faced a human); effects close to zero are drawn as dots.
\textbf{b}~Re-raises and responses across bid levels, human bids only. The left panel shows, for the reference cell, the share of human bids at each level that are re-raises or responses, with the share of rounds still active (green line; closing price at or above the level). The other panels show, for each level from 10 to 150, the effect on the probability that a human bid is a re-raise, from a linear probability model on the indicators of Eq.~(1) with categorical round controls; the response share changes by the same amount with the opposite sign. Dots mark levels with $p < 0.05$ (unadjusted). The green area (right axis) gives the share of rounds still active among the pairs in each panel. Higher levels are reached only in escalated rounds, so these comparisons describe the bids that occur there and are exploratory.
Bars in \textbf{a} and bands in \textbf{b} are 95\% confidence intervals, the estimate $\pm$ 1.96 standard errors from a wild bootstrap with one Rademacher weight per pair (for the median regressions in \textbf{a} following Feng et al., 2011), 2,000 draws. }
  \label{fig:raising}
\end{figure}

To see which kind of bid the labels act on, we count each participant's own bids per round, excluding the opening bid, split them into responses and re-raises, and estimate Eq.~(\ref{eq:main}) at the median for each (Fig.~\ref{fig:raising}a; exploratory). In the reference cell, the typical participant makes 4.2 bids per round after opening, 2.4 in response and 1.2 as re-raises; the medians do not add up, so the parts need not sum to the total. Told that a human was or might be the opponent, participants make 1.0 more bids per round (95\% CI [0.1, 1.8], $p = 0.03$), or 5 points of the price, and the increase lies in re-raises ($+0.6$, [0.1, 1.1], $p = 0.03$), not in responses (0.0, [$-0.2$, 0.2], $p = 0.92$). Told that an optimising machine might be the opponent, they make 1.0 fewer bids ($p = 0.06$), again through re-raises ($-0.7$, [$-1.4$, $-0.1$], $p = 0.03$) rather than responses ($-0.2$, $p = 0.13$). The two descriptions that move prices therefore act on the same margin in opposite directions: how often people raise again while already ahead.

Facing an agent changes the mix of bids rather than their amount. Against either agent, participants respond less often (AI-Imitate $-3.4$, [$-4.9$, $-2.0$]; AI-Optimal $-2.9$, [$-4.5$, $-1.3$]; both $p < 0.001$) and re-raise more ($+9.7$, [6.1, 13.2]; $+11.3$, [7.7, 14.9]; both $p < 0.001$), so that their own part of the price hardly changes ($+4.0$ and $+4.3$; $p = 0.16$ and $0.10$). The shift is concentrated at low bid levels, where the share of re-raises among human bids is 20 to 35 percentage points higher than against a human, and it fades beyond the prize (Fig.~\ref{fig:raising}b). Part of it may be mechanical: the agents take longer to bid (median 1.07~s against 0.68~s for human opponents), which leaves more time to raise again before they respond. Because participants' own part of the price does not fall, the lower prices against agents appear to arise on the agents' side of the bidding.

Across bid levels, the optimising-machine label lowers the share of re-raises at low levels, between 10 and 40, where most rounds are still active (Fig.~\ref{fig:raising}b). The human label leaves the share unchanged at every level while raising the number of re-raises, consistent with rounds that last longer and offer more occasions to raise again; the two analyses also weight participants differently, the decomposition each participant equally and the per-level shares each bid. The rise for the generic and imitating labels beyond a price of 120 concerns a few per cent of rounds, and we do not interpret it.

\subsection*{The label effects do not run through timing or beliefs about the opponent}

\begin{figure}[!htbp]
  \begin{minipage}[t]{0.49\linewidth}
    {\sffamily\bfseries\fontsize{8}{8}\selectfont a}\\[0.5mm]
    \includegraphics[width=\linewidth]{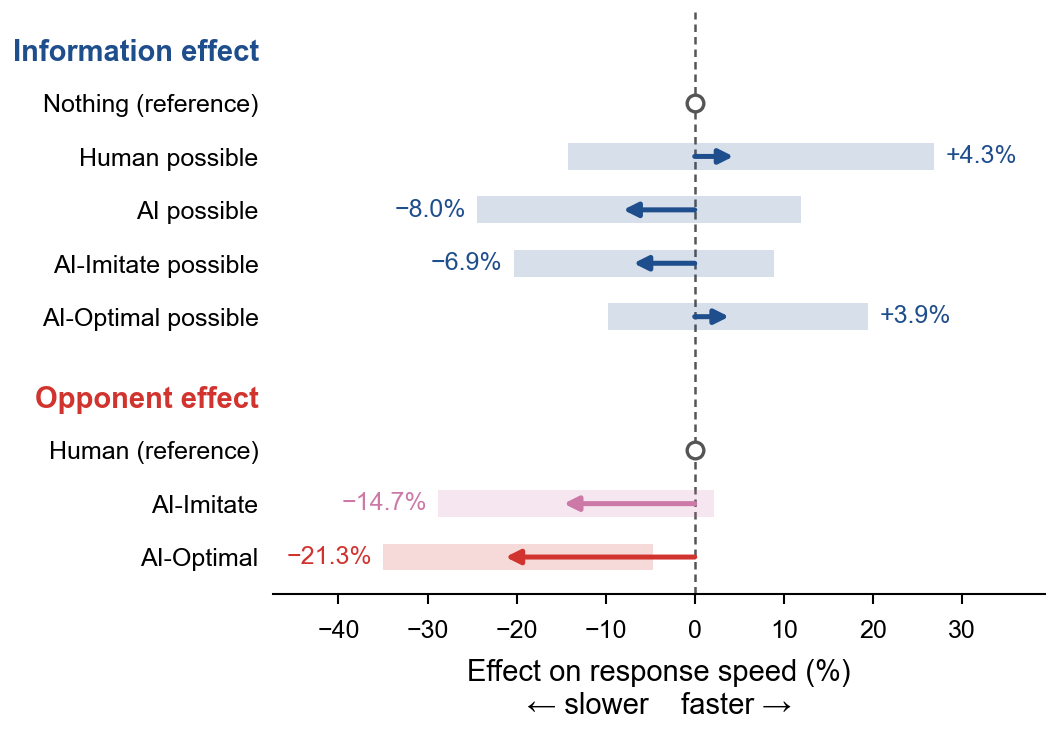}
  \end{minipage}\hfill
  \begin{minipage}[t]{0.49\linewidth}
    {\sffamily\bfseries\fontsize{8}{8}\selectfont b}\\[0.5mm]
    \includegraphics[width=\linewidth]{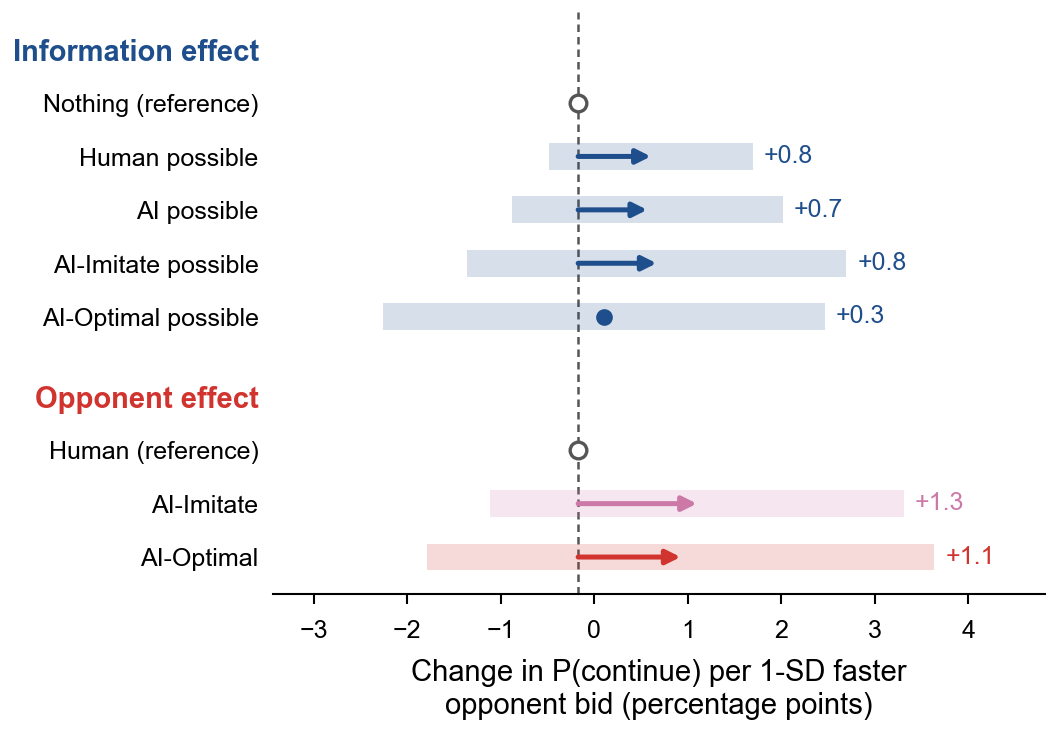}
  \end{minipage}
\caption{\textbf{What participants are told changes neither how fast they respond nor how they react to the opponent's pace, while facing an agent slows their responses.}
\textbf{a} Effects on own response speed at decision nodes at which the participant bid ($n = 116{,}316$ responses). A decision node is a point at which the opponent had just bid and the participant was behind; response time is measured from that bid. Estimates come from a regression of log response time on the indicators of Eq.~(1) with categorical controls for round and decision level (the standing bid, in 5-point steps, levels above 150 pooled), expressed as the percentage change in speed relative to the reference cell (told nothing, faced a human; positive values mean faster responses). Confidence intervals are computed on the log scale and then transformed.
\textbf{b} Responsiveness of continuation to the opponent's speed at all decision nodes ($n = 137{,}356$ from 1,395 participants in 1,033 pairs). Opponent speed is the standardised, sign-flipped time the opponent took for the bid that preceded the decision, capped at 10~s, so that higher values mean a faster bid. A linear probability model of continuing (bidding rather than conceding or timing out) includes opponent speed, the indicators of Eq.~(1), their interactions with opponent speed and the same categorical controls. Arrows start at the slope in the reference cell ($-0.2$ percentage points per standard deviation, 95\% CI [$-1.1$, 0.7]; dashed line) and end at the slope with one indicator switched on.
In \textbf{a} and \textbf{b}, shaded bars show the 95\% confidence interval of each effect, the estimate $\pm$ 1.96 standard errors from a wild cluster bootstrap by pair (Rademacher weights, 2,000 draws); a bar crossing the dashed line marks an effect not distinguishable from zero. AI-Imitate and AI-Optimal abbreviate the descriptions participants read. Source data are provided as a Source Data file.}
  \label{fig:effects}
\end{figure}

A description could change bidding by changing its pace. We measure each participant's response time at the decision nodes, the moments at which the opponent has just bid and the participant is behind (Fig.~\ref{fig:effects}a). No description changes how fast participants respond: told that a human was or might be the opponent, they respond 4.3\% faster (95\% CI [$-14.3$, 27.0], $p = 0.67$), and told that an optimising machine might be, 3.9\% faster ([$-9.8$, 19.4], $p = 0.61$); the generic and imitating labels are equally uninformative ($p = 0.41$ and $0.38$). The opponent matters instead. Against the agents, participants respond more slowly (AI-Imitate $-14.7\%$, [$-28.9$, 2.2], $p = 0.08$; AI-Optimal $-21.3\%$, [$-35.1$, $-4.7$], $p = 0.014$), in line with the agents' own slower pace.

A description could also change how people read the opponent's pace, for example as a signal of resolve when a person might be on the other side. We therefore relate the decision to continue at each decision node to the speed of the opponent's preceding bid (Fig.~\ref{fig:effects}b). In the reference cell, a faster opponent bid does not change the probability of continuing ($-0.2$ percentage points per standard deviation, [$-1.1$, 0.7]), and no description or opponent changes this slope (all $p > 0.16$). A specification with linear controls for bid level and round suggests a steeper slope under the human description, but this does not survive categorical controls, because continuation varies non-monotonically with the bid level.

The label effects also do not require believing that the opponent is human. After the final round, 78\% of participants guessed that they had faced an AI (1,091 of 1,395), including 67\% of those who had faced a human, while the agents were recognised as such by 90\% (Table~\ref{tab:guess_confusion}). Even among participants told for certain that their opponent was human, 61.4\% guessed AI (Supplementary Table~TODO). The human description thus raises prices although most participants end up believing they faced a machine. Because the guess was elicited after play, it cannot show what participants believed while bidding.

\begin{table}[!htbp]
\centering
\begin{threeparttable}
\caption{Post-game guess against the opponent actually faced}
\label{tab:guess_confusion}
\small
\begin{tabular}{lcccc}
\toprule
 & \textbf{Guessed human} & \textbf{Guessed AI} & \textbf{Total} & \textbf{Correct} \\
\midrule
Faced a human & 239 (33.0\%) & 485 (67.0\%) & 724 & 33.0\% \\
Faced AI-Imitate & 32 (9.5\%) & 304 (90.5\%) & 336 & 90.5\% \\
Faced AI-Optimal & 33 (9.9\%) & 302 (90.1\%) & 335 & 90.1\% \\
\midrule
Total & 304 & 1,091 & 1,395 & \\
Correct & 78.6\% & 55.5\% & & 60.6\% \\
\bottomrule
\end{tabular}
\begin{tablenotes}
\small
\item \textit{Notes:} Rows give the opponent actually faced, columns the incentivised guess after the final round. Cells give the number of participants and, in parentheses, the share of the row. \textit{Correct} in the last column is the share of the row whose guess matched the opponent; in the last row it is the share of each guess that was right, and in the bottom-right cell the share of all guesses that were right. Both members of a human--human pair guess, so these pairs contribute two participants.
\end{tablenotes}
\end{threeparttable}
\end{table}

\begin{figure*}[!htbp]
    \centering
    \includegraphics[width=\textwidth]{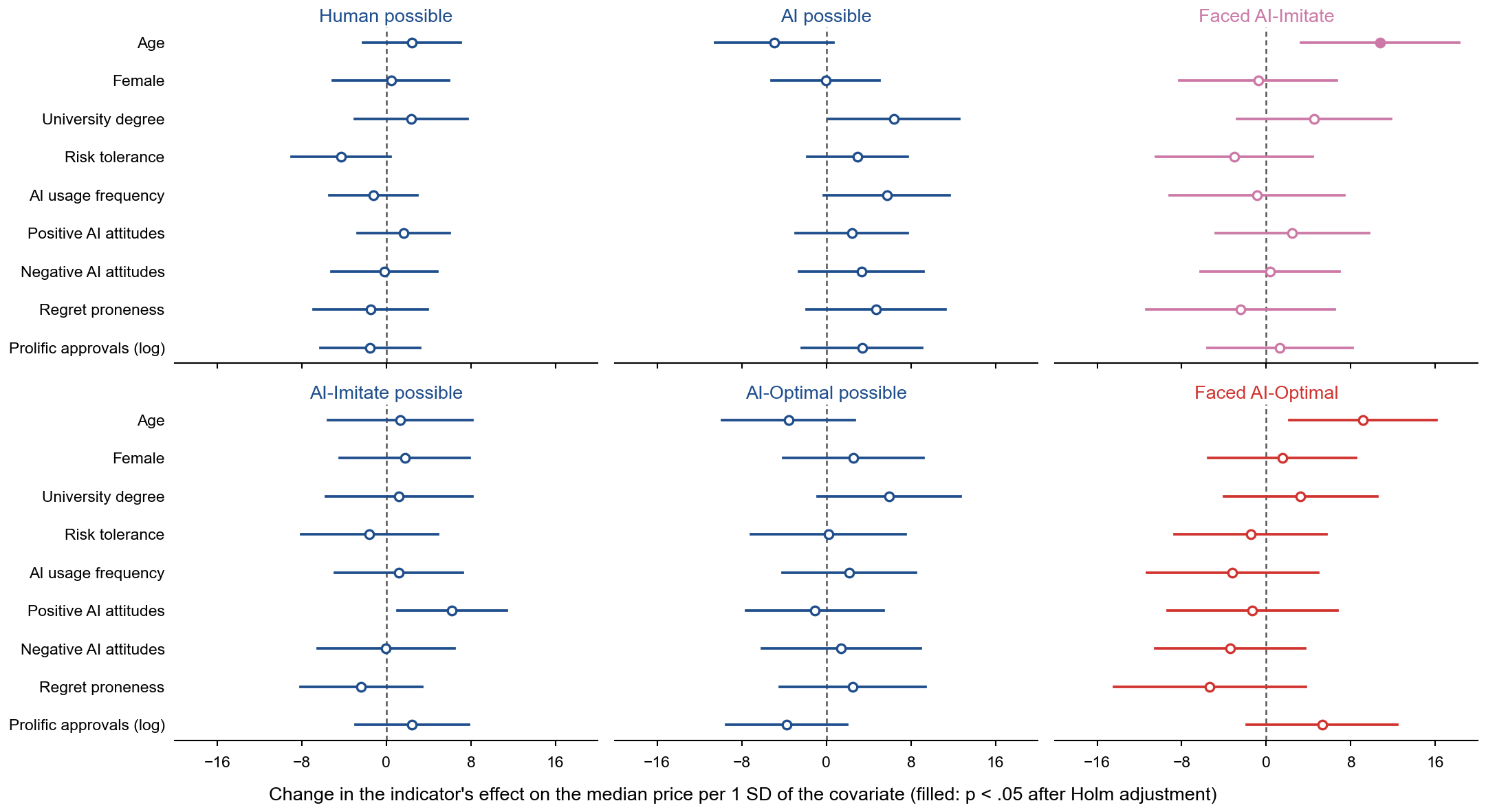}
    \caption{\textbf{The information effects do not vary detectably with participant characteristics, while the opponent effect is weaker among older participants.}
    Each panel refers to one indicator of Eq.~(1); rows are participant characteristics. Dots show how much the indicator's effect on the median pair-level average closing price changes per one standard deviation of the characteristic, from median regressions ($\tau = 0.5$) of Eq.~(1) extended by the characteristic and its interactions with all six indicators, one characteristic per model. A dot at zero means that the indicator's effect does not depend on the characteristic. Characteristics are averaged within human--human pairs and standardised across pairs: age, female, university degree, risk tolerance (0--10), frequency of AI use, positive and negative attitudes towards AI (GAAIS-10), regret proneness and the number of prior Prolific approvals (log). Bars are 95\% confidence intervals, the estimate $\pm$ 1.96 standard errors from a wild bootstrap with one Rademacher weight per pair \citep{feng2011wild}, 2,000 draws, unless stated otherwise. Filled dots mark interactions with $p < 0.05$ after Holm adjustment within each indicator across the nine characteristics; hollow dots are not significant after adjustment. $n = 1{,}033$ pairs. This analysis is exploratory.}
    \label{fig:interactions}
\end{figure*}

Finally, the label effects do not differ detectably across participants. We interacted each indicator of Eq.~(\ref{eq:main}) with nine characteristics recorded before participants read the opponent description: age, gender, education, risk tolerance, frequency of AI use, positive and negative attitudes towards AI, regret proneness and prior experience on Prolific. No interaction with a description survives a Holm adjustment within each indicator (Fig.~\ref{fig:interactions}). The only robust moderation concerns the opponent: the price reduction against AI-Imitate is concentrated among younger participants (10.8 per standard deviation of age, Holm-adjusted $p = 0.04$). These analyses are exploratory.

\section*{Discussion}

Separating what people are told about an opponent from whom they face shows two distinct influences on competitive escalation. Holding the opponent fixed, describing it as a possible human raised the typical price, and describing it as a possible optimising machine lowered it, with the two descriptions about 15\% of the prize value apart. The opponent itself mattered as well: facing an agent also lowered prices and, in addition, made it rarer that both players finished with positive earnings.

These results add an informational determinant to research on escalation of commitment, which has traced escalation mainly to features of the decision-maker and the decision \citep{staw1976knee, brockner1992escalation, sleesman2012cleaning}. Unlike most of those, the description of an opponent is set by whoever runs the contest. The bidding data also show where escalation arises. Much of it comes not from refusing to concede when behind but from raising again while already ahead, which increases one's own payment without any move by the opponent forcing it. The prize itself is crossed mainly by responding, as in the classic dollar-auction trap, and raising while ahead resumes beyond it. In our exploratory analyses, the two descriptions that move prices act on this margin in opposite directions, while leaving responses when behind unchanged.

Why a possible human increases and an optimising machine reduces raising while ahead cannot be settled with our data, but some explanations can be ruled out. The effects do not require believing that the opponent is human: most participants ended up guessing that they had faced an AI, including 61.4\% of those told for certain that their opponent was human. They also do not run through timing, since neither participants' own pace nor their responsiveness to the opponent's pace changed with the description, and they were not concentrated among participants with particular characteristics. What remains is consistent with a framing effect in the sense of \cite{liberman2004name}, in which the description changes what participants expect from the contest without changing the game. A possible person may engage a competitive motive that shows in pushing while ahead, whereas a machine described as playing optimally may make such pushing look futile. That neither a generic AI label nor an AI trained to imitate humans had a detectable effect fits this reading: the content of the description carries the effect, not the distinction between human and machine as such.

The opponent effect, by contrast, describes the agents we built. Pairs with an agent rarely reached the low-price truce in which both players finish ahead, and both agents learned from human pairs that rarely reached it themselves. Machines sustain cooperation when they are built for it \citep{crandall2018cooperating}; ours were not. Against our agents, participants also shifted from responding to raising again while ahead, which may partly reflect the agents' slower pace. The information effect is estimated net of these features and is therefore not a by-product of how our agents play.

For disclosure rules, the content of a disclosure is a behavioural choice. The EU AI Act requires that people be told when they interact with an AI system but leaves open what they should be told, and it exempts cases in which AI involvement is obvious \citep{euaiact2024}. Our participants recognised the agents in about 90\% of cases, yet the descriptions still shifted escalation by an economically meaningful amount, so the inferability of an AI does not make an explicit description behaviourally redundant. Whether a description that lowers or raises escalation is desirable depends on the setting; our results show that the choice between them is not neutral.

Several limits bound these results. 
Treatments were assigned by session, so effects of what participants were told could in principle reflect other differences between sessions.
The contest is stylised: an online all-pay auction reproduces the incentives behind escalation but lacks the relationships and exit options of real competition, so how the effect sizes carry over is open. The sample is a UK online panel, younger and more digitally experienced than the general population. Our two agents converged to similar play against humans, so the opponent effect should be read as the effect of facing a non-human opponent rather than of a particular strategy. Beliefs were measured once, after the final round, and the analyses of bidding behaviour are exploratory; effects at prices beyond the prize rest on the minority of rounds that reach them.

Eliciting beliefs during play would show whether a description moves expectations before behaviour diverges. Physiological measures of arousal, in a design that separates description and opponent as ours does, could show what a possible human engages when the behavioural timing measures do not change. A field setting with real stakes, such as a procurement platform that varies the disclosure shown to human bidders facing algorithmic bids, would test how far the effects carry. 

Describing an AI competitor is not a neutral act. What people are told about whom they face changes how they compete, separately from what the opponent does, and the direction depends on the wording. An organisation that treats such a disclosure as a compliance step makes a choice about behaviour without recognising it as one.

\section*{Methods}

\subsection*{Participants}
We recruited participants via Prolific between February and March 2026, restricted to UK residents to hold language and currency constant. One participant was excluded for evidence of automated responding, leaving 1{,}395 participants in 1{,}033 pairs; human--human pairs contain two participants and pairs with an AI agent contain one. 
Data were collected in twelve sessions between 10 February and 22 March 2026. The session in which participants were told for certain that they faced a human ran first, with 206 pairs, because its play provided the training data for the AI agents. For practical reasons, the remaining sessions then ran in two blocks: first those in which AI-Optimal could not be faced (28 February to 2 March), then those in which it could (19 to 22 March).

Before starting, participants read a description of the task, the data recorded, and their right to withdraw, and indicated consent before proceeding. The study was approved by the KIT IRB (A2025-082) and preregistered at AsPredicted (\url{https://aspredicted.org/a8cd87.pdf}) with amendments outlined in the Supplementary Information Section 2.

The median completion time was 17 minutes. Participants received a fixed fee of \pounds2.10 regardless of performance and a bonus of \pounds0.01 per 5 points, paid only on a positive final balance (limited liability), giving an effective rate of approximately \pounds11.50 per hour.

Following \cite{rilla2026recognising}, entry required a CAPTCHA, and the comprehension questions and the post-auction questionnaire contained honeypot questions. Participants also had to answer comprehension questions on the scoring rule correctly before the auction began.

\subsection*{Task}
Two players compete for a fixed prize of 100 points. Each sees the current highest bid and their own bid, updated in real time. A click raises the participant's own bid to 5 points above the current highest, and either player may click at any moment, including the player who is already ahead. Bids are not capped.

A round opens with both bids at zero and can end once at least one player has bid. From that point, the player whose bid is lower may end the round, and the round also ends automatically if that player neither bids nor concedes within 10 seconds. The same rule governs the AI agents. When a round ends, both players pay their own final bid, and the higher bidder receives the prize. A participant who bids 60 and wins earns 40; one who bids 55 and loses pays 55.

Each pair plays 30 rounds against the same opponent. Points accumulate across rounds, and after each round participants see the round number, the result of the round, and their running balance. Participants cannot lose money: a negative final balance yields no bonus rather than a deduction (limited liability).

Before the auction, participants read a walkthrough of the rules with worked examples and answered comprehension questions on the scoring rule. There were no practice rounds, since playing one would have meant interacting with the opponent before the auction began. The interface is identical in every condition.

\subsection*{Design}
Each participant read information in the instructions describing the opponent and then played against another Prolific worker, AI-Imitate, or AI-Optimal. The information varies in two respects. The first is certainty: one information box names one opponent, gives an even chance between two, or names only ``an opponent''. The second is how an AI is described, either by what it was trained to do, as ``an AI trained to imitate humans'' or ``an AI trained to play optimally'', or by the bare label ``an AI''. The full texts of the information boxes are reproduced in Supplementary Section~\ref{si:instructions}.

Every statement was true and every stated probability genuine, which constrains the crossing of description and opponent. A participant told the opponent was human faced a human; one told ``an AI'' faced either agent; one told there was an even chance between two types was assigned to one of them with probability one half. 
Treatments were assigned at the level of the session. A treatment combines a description of the opponent with the opponent types it could be paired with, and each treatment ran as its own session on Prolific under an identical study listing, with the same eligibility criteria and payment. Participants could not see their treatment before taking part, and each Prolific account could take part only once. Session-level assignment is common in experimental economics \citep{frechette2012session}. In sessions whose description named an even chance between two opponent types, the opponent was drawn at random on arrival; in the others, the assigned treatment determined it. Some descriptions therefore span several sessions: ``an opponent'' ran as three sessions, one for each opponent type, and ``human or AI'' as two, one for each agent. Combinations of description and opponent that would require a false statement do not occur (Table~1). Both effects remain identified: each opponent type occurs under several descriptions, and most descriptions occur with more than one opponent type. Pairs told only ``an opponent'' that faced a human form the reference cell throughout. 

Participants who were to face a human were matched by arrival on Prolific, and both members of a pair read the same information, so the information condition is assigned at the level of the pair for human--human pairs and at the level of the participant for pairs with an agent. 

The treatment indicators record which opponents a participant was told were possible, one each for a human, an AI described as imitating, an AI described as optimising, and an AI named without further description. An indicator is active whenever its type was named as possible, whether for certain or with an even chance. A participant told there was an even chance between a human and an imitating AI therefore has two indicators active, and a participant told only that they face ``an opponent'' has none. The experiment was implemented in oTree 5.10 \citep{chen2016otree}.

\subsection*{AI agents}
Both agents share a two-headed feedforward neural network trained on the $\approx$104{,}000 decision actions from the 206 human pairs in the cell naming a human for certain. The network observes the game-state features available to a human player: bid level, which player leads, recent bidding activity, round number, and session progress. 
One head predicts the action, whether to continue bidding or to concede: the agent concedes when the predicted probability of conceding exceeds a threshold calibrated on the training data for each bidding deficit, and only when it is at least two steps behind. The second head predicts a response-time distribution, and the agent acts after its median, with a small random jitter, so that it bids on a human timescale.
Because the agents never concede when only one step behind, a participant can win a round against an agent only by raising again while ahead (Supplementary Section~S1.7). 

AI-Imitate is trained by behaviour cloning to reproduce typical human play. AI-Optimal applies advantage-weighted regression to the same data, weighting each decision by the round's outcome relative to the average outcome in similar states, so that decisions leading to above-average earnings receive greater weight; the resulting policy is regularised towards AI-Imitate by a KL penalty to keep it behaviourally plausible. 

Against human participants across all treatments, AI-Optimal earned more than AI-Imitate over the 30 rounds (mean $+171$ against $-12$ points; medians $-85$ and $-102$). Rounds won is not the relevant criterion, since a round won at a price above the prize is worse than one conceded early: AI-Imitate won $51.3\%$ of rounds and AI-Optimal $49.9\%$. Both agents are deployed with fixed parameters and do not adapt to the participant, so every participant in a given condition met the same model. We do not claim that AI-Optimal plays a game-theoretically optimal strategy; we are not aware of a characterised equilibrium for a repeated dynamic all-pay auction of this form, and the phrase shown to participants describes the training objective.

\subsection*{Measures}
The primary outcome is the average closing price across the 30 rounds, taken at the level of the pair, where the closing price of a round is its final standing bid.

Secondary outcomes are mutual success, defined as both players in a pair ending the 30 rounds with positive cumulative points, and, at the level of the individual decision, whether a participant continued bidding. In pairs with an agent, mutual success requires a positive balance for both the participant and the agent. A decision node is a point at which the participant was behind and the opponent had just raised. Because either player may raise at any moment, raises following the participant's own previous action are not decision nodes and are excluded; if the opponent raised twice in a row, the node follows the second raise. Continuation is coded 1 where the participant bid and 0 where the round ended by concession or by timeout.

Every bid after the opening bid of a round is either a response, a bid that follows the opponent's bid, or a re-raise, a bid that follows the bidder's own bid while they already lead. A participant's own bids per round are the number of bids they placed in a round, excluding the opening bid, split into responses and re-raises and averaged over the 30 rounds, with rounds without an own bid counting as zero. Because each bid raises the price by 5, five times this number is the part of the price a participant added through their own bids.

Opponent response speed is built from the time the opponent took to make the raise that placed the participant behind, measured from the preceding action and capped at 10~s. It is standardised and sign-flipped, so that higher values correspond to faster raises. The distribution is left-skewed. Own response time is the time a participant took to bid at a decision node, measured from the opponent's raise.

Before the auction, and before reading the opponent description, participants completed a questionnaire recording age, gender, education, risk preference on a 0--10 scale \citep{dohmen2011individual}, attitudes towards AI (GAAIS-10, \citealp{schepman2026validation}), regret proneness \citep{schwartz2002maximizing}, AI usage frequency, and prior Prolific approvals. For the heterogeneity analysis, AI usage frequency is coded as an ordinal score from 0 (never) to 5 (daily) and approvals enter on the log scale. After the final round, they guessed whether the opponent had been human or AI, incentivised at \pounds0.15 for a correct guess; and answered three open-text questions on their strategy, the reasoning behind their guess, and the study as a whole, which are not analysed here.

\subsection*{Statistical analysis}

All treatment effects are estimated from Eq.~(\ref{eq:main}), with told nothing and faced a human as the omitted categories, so that the intercept describes the reference cell. The primary outcome, the pair-level average closing price, and the secondary outcomes, mutual success and continuation at decision nodes, were preregistered. The inference procedure was not specified in the preregistration; it is described below and applied uniformly. All analyses of how bidding unfolds within rounds, of response times, of beliefs and of heterogeneity are exploratory.

\textit{Prices.} Pair-level average prices are estimated by quantile regression at $\tau = 0.5$, because the price distribution has a heavy right tail and a conditional mean would track the escalated minority rather than the typical pair. The contrast between the human and the optimising-machine descriptions is the difference of the two coefficients, computed within each bootstrap draw. Per-round effects estimate the same model separately for each round, with one observation per pair. Price trajectories by opponent are per-round medians, with intervals from a bootstrap that resamples pairs together with their full 30-round trajectories.

\textit{Mutual success.} Mutual success is estimated by logistic regression at the level of the pair. Effects are reported as average marginal effects, computed as the average change in predicted probability when an indicator switches from 0 to 1, and the reference cell as its predicted probability.

\textit{Decision-level models.} Continuation at decision nodes is estimated by a linear probability model with opponent speed, the indicators of Eq.~(\ref{eq:main}), their interactions with opponent speed and categorical controls for round and for the decision level, the standing bid in 5-point steps with levels above 150 pooled. Own response speed is estimated on the decision nodes at which the participant raised, by ordinary least squares of the log response time on the indicators with the same categorical controls; response times below 0.05~s are set to 0.05~s, and coefficients are reported as the percentage change in speed, $100(e^{-\beta} - 1)$. A specification with linear controls for round and decision level is reported in Supplementary Table~\ref{sitab:speed_linear}.

\textit{Responses and re-raises.} The number of own bids per round and its parts from responses and from re-raises (Measures) are each estimated by quantile regression at $\tau = 0.5$ on the indicators of Eq.~(\ref{eq:main}), with one observation per participant. Because medians are not additive, the effects on the two parts need not sum to the effect on the total.

\textit{Heterogeneity.} For each of nine participant characteristics recorded before the opponent description was shown, Eq.~(\ref{eq:main}) for the pair-level price is extended by the characteristic and its interactions with all six indicators, one characteristic per model. Characteristics are averaged within human--human pairs and standardised across pairs. Within each indicator, the nine interaction $p$-values are adjusted with Holm's step-down procedure.

\textit{Beliefs.} The post-game guess is analysed descriptively; shares are reported with intervals from a bootstrap that resamples pairs, because both members of a human--human pair guess.

\textit{Inference.} All intervals and $p$-values come from bootstrap procedures with 2,000 draws that treat the pair as the independent unit. For quantile regressions we use the wild bootstrap of \cite{feng2011wild}, which re-estimates the model on outcomes formed by adding the absolute residual, multiplied by a Rademacher weight drawn once per pair, to the fitted value. This avoids estimating the density of the outcome at the median, on which analytical standard errors for quantile regression depend and which is poorly determined here because prices are heavy-tailed and cell sizes range from 27 to 206 pairs. For decision-level linear models we use the wild cluster bootstrap with Rademacher weights clustered by pair \citep{davidson2008wild}. For the logistic regression and for descriptive shares we resample pairs. Confidence intervals for effects are the estimate $\pm$ 1.96 bootstrap standard errors, and $p$-values are two-sided from the same standard errors; descriptive quantities such as shares and per-round medians are reported with percentile intervals. Random seeds are fixed, and analyses that share a model use the same draws.

\textit{Software.} Analyses were conducted in Python with statsmodels, NumPy and pandas.

\section*{Acknowledgments}

For helpful comments and discussions, we thank Holger Herz, Adam Zylbersztejn, and seminar participants at the 4th Workshop on Gender in Adaptive Design (KIT), the KD\textsuperscript{2}School Retreat, the BSE Summer Forum Workshop on Computational and Experimental Economics, the Berlin Symposium on AI and the Future of Work, the TIBER symposium and the European Meeting of the Economic Science Association (ESA) in Barcelona. Mika Baczweski and Paul Althaus provided excellent research assistance. This research was funded by the Deutsche Forschungsgemeinschaft (DFG, German Research Foundation) -- GRK2739/2 -- Project Nr.\ 447089431 -- Research Training Group KD\textsuperscript{2}School: Designing Biosignal-Adaptive Systems for Decision-Making Processes, and by the Joachim Herz Stiftung.

\section*{Data and Code}

All data and code will be made publicly available on AsPredicted's ResearchBox upon publication.

\bibliographystyle{plainnat}
\bibliography{references}

\end{document}